\documentclass{article}
\usepackage{spconf}
\usepackage{float}
\usepackage{amsmath}
\usepackage{booktabs}
\usepackage{graphicx}
\usepackage{enumitem}
\usepackage{epsfig}
\usepackage{subcaption}
\usepackage{amssymb}

\title{Audio-Linguistic Embeddings for Spoken Sentences}

\name{Albert Haque$^1$  \qquad Michelle Guo$^1$  \qquad Prateek Verma$^2$  \qquad Li Fei-Fei$^1$ }

\address{
$^1$Department of Computer Science, Stanford University \\
$^2$Center for Computer Research in Music and Acoustics, Stanford University
}

\begin{document}

\maketitle

\begin{abstract}
We propose spoken sentence embeddings which capture both acoustic and linguistic content.
While existing works operate at the character, phoneme, or word level, our method learns long-term dependencies by modeling speech at the sentence level.
Formulated as an audio-linguistic multitask learning problem, our encoder-decoder model simultaneously reconstructs acoustic and natural language features from audio.
Our results show that spoken sentence embeddings outperform phoneme and word-level baselines on speech recognition and emotion recognition tasks.
Ablation studies show that our embeddings can better model high-level acoustic concepts while retaining linguistic content.
Overall, our work illustrates the viability of generic, multi-modal sentence embeddings for spoken language understanding. 
\end{abstract}

\section{Introduction}\label{sec:intro}

During verbal communication, humans process multiple words sequentially, often waiting for a full sentence to be completed.
Yet, many written and spoken language systems depend on individual character and word-level representations either implicitly or explicitly.
This lack of sentence-level context can make it difficult to understand sentences containing conjunctions \cite{kitaev2018constituency}, negations \cite{li2018learning}, and vocal pitch variations \cite{mozziconacci1995pitch}.
In this work, we investigate embeddings in the domain of spoken language processing and propose spoken sentence embeddings, capable of modeling both acoustic and linguistic content in a single latent code.

Machine representation of words dates back to ASCII.
This one-hot representation encodes each character using a mixture of dummy or indicator variables.
While this was slowly extended to words, the large vocabulary size of languages made it difficult.
Learned, or distributed, word vector representations \cite{word2vec} replaced one-hot encodings.
These word vectors are able to capture semantic information including context from neighboring words.
Even today, the community continues to build better contextual word embeddings such as ELMo \cite{peters2018deep}, ULMFit \cite{howard2018universal}, and BERT \cite{devlin2018bert}.
Word, phoneme, and grapheme embeddings like Speech2Vec \cite{chung2018speech2vec} and Char2Wav \cite{sotelo2017char2wav} have also been proposed for speceh, following techniques from natural language understanding.

While word-level embeddings are promising, they are often insufficient for speech-related tasks for several reasons.
First, word and phoneme embeddings capture a narrow temporal context, often a few hundred milliseconds at most.
As a result, these embeddings cannot capture long-term dependencies required for higher-level reasoning (e.g., paragraph or song-level understanding).
Almost all of the systems for speech recognition focus on the correctness of local context (e.g., letters, words, and phonemes) rather than overall semantics.
Second, for speech recognition, an external language model is often used to correct character and word-level predictions.
This requires the addition of complex, multiple hypothesis generation methods \cite{deepspeech2}. 

Sentence-level embeddings offer advantages over word and character embeddings.
A sentence-level embedding can capture latent factors across words.
This is directly useful for higher-level audio tasks such as emotion recognition, prosody modeling, and musical style analysis.
Furthermore, most external language models operate at the sentence-level.
By having a single sentence-level embedding, the embedding can capture both acoustic and linguistic content at longer contextual window sizes -- thus alleviating the need for an external language model entirely, by learning the temporal structure.

\textbf{Contributions.} In this work, our contributions are two-fold.
First, we propose moving from phoneme, character, and word-level representations to sentence-level understanding by learning spoken sentence embeddings.
Second, we design this embedding to capture both linguistic and acoustic content in order to learn latent codes which can be applicable to a variety of speech and language tasks.
We verify the quality of the embedding in our ablation studies, where we assess the generality of sentence-level embeddings when used for automatic speech recognition and emotion classification.
We believe this work will inspire future work in speech processing, semantic understanding, and multi-modal transfer learning.

\section{Method}

Our method allows us to learn spoken sentence embeddings that capture both acoustic and linguistic content.
In this section, we discuss (i) how we handle long sequences with a temporal convolutional network \cite{bai2018empirical} and (ii) how to learn audio-linguistic content under a multitask learning framework \cite{caruana1997multitask}.

\subsection{Temporal Convolutional Network (TCN)}

Our goal is to learn a spoken sentence embedding, which can be used for a variety of speech tasks. 
Recurrent models are often the default starting point for sequence modeling tasks \cite{bai2018empirical}.
For most applications, the state-of-the-art approach to start with is very often a recurrent model.
This is evident in machine translation \cite{sutskever2014sequence, bahdanau2014neural}, automatic speech recognition \cite{chan2016listen, graves2013speech, graves2014towards}, and speech synthesis tasks \cite{wang2017tacotron, shen2018natural, sotelo2017char2wav}.

\begin{figure}[t]
    \centering
    \includegraphics[width=1.0\linewidth]{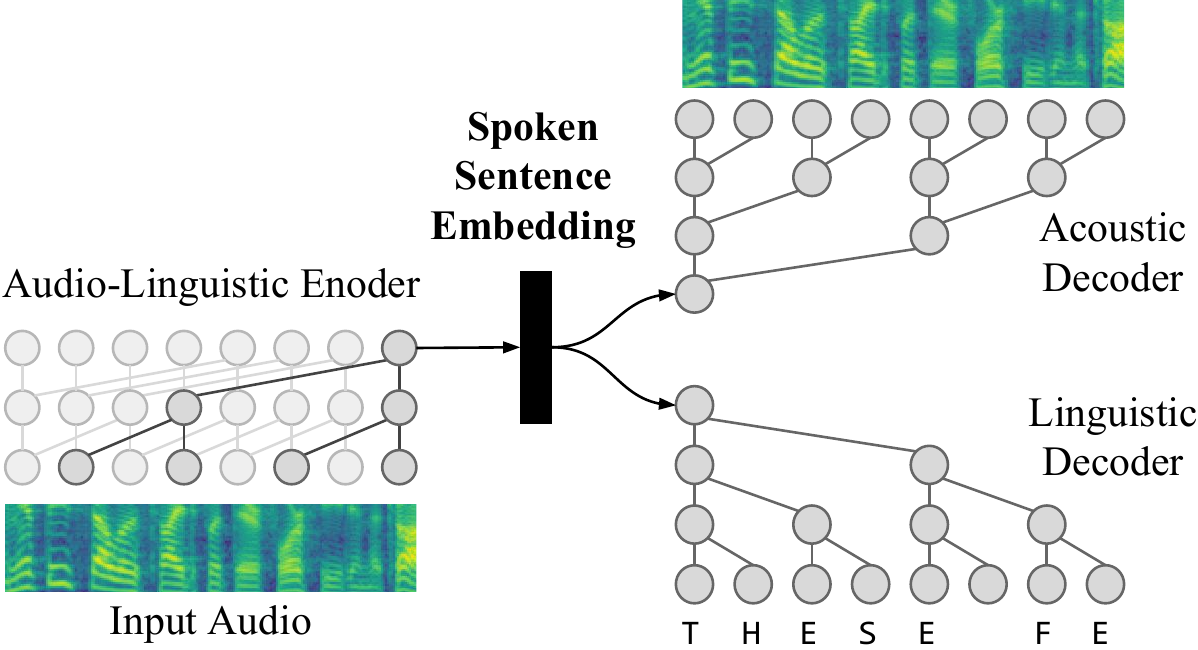}
    \caption{\textbf{Audio-linguistic embedding for spoken sentences.} Formulated as a multitask learning problem, our method learns a multi-modal spoken sentence embeddings by reconstructing linguistic and acoustic features during training.}
    \label{fig:method}
\end{figure}

However, recurrent models such as recurrent neural networks (RNNs) are notoriously difficult to train \cite{pascanu2013difficulty}.
For years, machine learning researchers have tried to make RNNs easier to train through novel training strategies \cite{bengio2015scheduled} and architectures \cite{koutnik2014clockwork}.
In \cite{bai2018empirical}, the authors show that fully convolutional networks can outperform recurrent networks, without the training complexities \cite{bai2018empirical}.
In addition, they can better capture long term dependencies \cite{oord2016wavenet} required for such tasks. 
Motivated by these findings, in this work, we opt for a fully convolutional sequence model (Figure \ref{fig:method}).
Similar to WaveNet \cite{oord2016wavenet}, we use a temporal convolutional network (TCN) \cite{bai2018empirical}.
While we use the TCN in this work, any causal model will suffice (e.g., Transformer \cite{vaswani2017attention}).

\textbf{Causal Convolutions.}
To begin, we introduce some quick notation.
The sequence modeling task is defined as follows.
Given an input sequence $x$ of length $T$, we have $x = x_1, ..., x_T$, where each $x_t$ is an observation for timestep $t$.
Suppose we wish to make a prediction $y_t$ at each timestep, then we have $y=y_1, ..., y_T$.
The \textit{causal constraint} states that when predicting $y_t$, it should depend only on past observations $x_{<t}$ and not future observations.
For example, a bidirectional RNN does not satisfy this constraint.

\textbf{Dilated Convolutions.}
Standard convolutions have a fixed filter size and thus have a fixed temporal understanding.
If our goal is to learn sentence embeddings, we need filters capable of modeling longer temporal windows -- ideally modeling the entire sentence.
Following the work of \cite{oord2016wavenet, bai2018empirical}, we employ dilated convolutions that enable an exponentially large temporal context window at different layers of the TCN.
For an input sequence $x=x_1,...,x_T\in \mathbb{R}^T$ of length $T$ and a filter $f:\{0, ..., k-1\} \rightarrow \mathbb{R}$ for some filter size $k$, the dilated convolution operation $F$ on element $s$ of the sequence is:
\begin{equation}
    F(s) = (x \circledast_d f)(s) = \sum\limits_{i=0}^{k-1} f(i) \cdot x_{<(T-di)}
\end{equation}
where $d$ is the dilation factor and $\circledast$ denotes the convolution operator.
When $d=1$, a dilated convolution reduces to a standard convolution.
Figure \ref{fig:method} shows a TCN with dilation factor $d=2$ and kernel size $k=2$.
The TCN allows us to model entire sentences, typically with hundreds to thousands of timesteps, with a single encoder.
This encoder produces a fixed-size embedding vector which contains both audio and linguistic information from the input.
We explain this process more in the next section.

\subsection{Multitask Learning of Acoustics and Linguistics}\label{sec:multitask}

In the previous section, we explained how to convert a variable-length spoken sentence into a single, fixed-length vector using a TCN.
We now discuss how to encode both acoustic and linguistic content into this vector using multitask learning. 

A \textit{task} is a specific prediction problem (e.g., speech recognition, emotion recognition, speaker verification).
\textit{Multitask} learning attempts to train a model to simultaneously perform multiple tasks.
The simplest method is to combine the task-specific loss functions $\mathcal{L}_k$ into a single loss criterion $\mathcal{L}_{\textrm{Total}} = \sum_{k=1}^{K} \lambda_k \mathcal{L}_k$, where $\lambda_k$ is an arbitrary task-weighting coefficient, where $k$ refers to a task  index.
We use hard parameter sharing \cite{ruder2017overview} as the TCN encoder is shared for both the acoustic and linguistic decoders.
The acoustic target is the original mel-spectrogram input, whereas the linguistic target is a text transcription.

\section{Experiments}

\begin{table*}[t]
\centering
\caption{\textbf{Comparison of phoneme, word, and sentence embeddings.} Sentence embeddings are our proposed method. Phoneme and word error rates are denoted as PER and WER, respectively. All numbers denote percentages. Rows 5 and 6 do not have fusion because the model produces a sentence embedding. Plus/minus denotes 95\% confidence interval.}
\label{tab:sota}
\begin{tabular}{@{}lll|cc|ccc@{}}
\toprule
 & &  & \multicolumn{2}{c}{Speech Recognition} & \multicolumn{3}{|c}{Emotion Recognition} \\ 
\# & Embedding Level & Fusion Method & TIMIT PER & LibriSpeech WER & Accuracy & Precision & Recall \\ \midrule
1 & Phoneme & Average \cite{cer2018universal} &  $50.4 \pm 6.0$ & $74.8 \pm 9.6$ & $20.8 \pm 5.4$ & $19.5 \pm 1.8$ & $19.9 \pm 1.9$ \\
2 & Phoneme & Deep Avg Network \cite{iyyer2015deep} &  $50.8 \pm 6.5$ & $ 70.5 \pm 9.3 $ & $21.1 \pm 5.0$ & $17.5 \pm 6.0$ & $21.2 \pm 5.0$ \\
3 & Word \cite{word2vec,chung2018speech2vec} & Average \cite{cer2018universal} & $49.6 \pm 7.0$ & $ 30.8 \pm 6.0 $ & $21.9 \pm 3.0$ & $20.3 \pm 2.3$ & $21.5 \pm 4.0$ \\
4 & Word \cite{word2vec,chung2018speech2vec} & Deep Avg Network \cite{iyyer2015deep} & 51.2 $\pm$ 5.8  & $ 18.5 \pm 3.0 $ & $21.2 \pm 4.3$ & $23.0 \pm 6.6$ & $23.6 \pm 3.7$ \\ \midrule
5 &  \multicolumn{2}{l|}{Sentence - RNN (LSTM) \cite{hochreiter1997long}}  & \textbf{29.6 $\pm$ 4.0} &  \textbf{10.8 $\pm$ 1.8}   & $25.7 \pm 5.5$ & $21.0 \pm 5.0$ & $24.0 \pm 6.0$ \\
6 & \multicolumn{2}{l|}{Sentence - Fully Convolutional \cite{bai2018empirical}} 
 & $30.1 \pm 4.5$ & $ 13.5 \pm 2.0 $ & \textbf{29.2 $\pm$ 3.1 }& \textbf{28.8 $\pm$ 5.4} & \textbf{29.2 $\pm$ 3.1} \\ \bottomrule
\end{tabular}
\end{table*}

Our experimental procedure consists of two parts.
First, we learn individual phoneme, word, and sentence embeddings from TIMIT and LibriSpeech.
These include previously published baselines and our proposed spoken sentence embeddings.
Second, we evaluate these embeddings on automatic speech recognition and emotion recognition.

\subsection{Configuration}

\textbf{Datasets.} We train our model on the LibriSpeech dataset \cite{panayotov2015librispeech}.
The training set consists of 460 hours of 16 kHz English speech.
For speech recognition, we use both LibriSpeech and the TIMIT dataset \cite{timit} and report results on the test set.
For emotion recognition, we use the Ryerson Audio-Visual Database of Emotional Speech and Song (RAVDESS) dataset \cite{livingstone2018ryerson}.
It consists of 24 speakers demonstrating 7 emotions.
Because no official train-test split is provided, results are reported using four-fold cross-validation (75\% train, 25\% test). The input representation used in all methods are log-mel spectrograms with 80 mel filters, with audio sampled at 16kHz.

\textbf{Evaluation Metrics.}
Speech recognition performance is evaluated using the phoneme (PER) and word error rate (WER).
Emotion recognition, a multi-class classification problem, is evaluated using accuracy, precision, and recall. 

\subsection{Baselines}
We now discuss the baselines we compare our method against.
To compute a sentence-level embedding, one must decide (i) which intermediate embedding to use and (ii) how to fuse them into a single, sentence-level embedding.

\subsubsection{Intermediate Embedding}
We select two baseline embeddings for learning intermediate speech representations.
The embeddings were selected due to their ease of use and applicability to language tasks.
\begin{enumerate}[itemsep=0pt,topsep=4pt,leftmargin=12pt]
    \item \textit{Speech2Vec} \cite{chung2018speech2vec}. The Speech2Vec method learns an embedding for spoken words. Word alignments (i.e., segmentations) are required, but the method can be trained in an unsupervised manner. Different utterances of the same word will have different embeddings. This is the same as Word2Vec \cite{word2vec}, but instead applied to spoken language.
    \item \textit{Phoneme2Vec}. This is the same as Speech2Vec but applied to phonemes. Each utterance of a phoneme is encoded as a single embedding. This is useful for fine-grained tasks.
\end{enumerate}

\noindent
We train Speech2Vec and Phoneme2Vec on LibriSpeech, similar to the procedure in \cite{chung2018speech2vec}.
The word and phoneme alignments are computed using a Gaussian mixture model \cite{mcauliffe2017montreal}.

\begin{figure*}[t]
\vspace{-3mm}
    \centering
    \begin{subfigure}[b]{0.33\textwidth}
        \includegraphics[width=\textwidth]{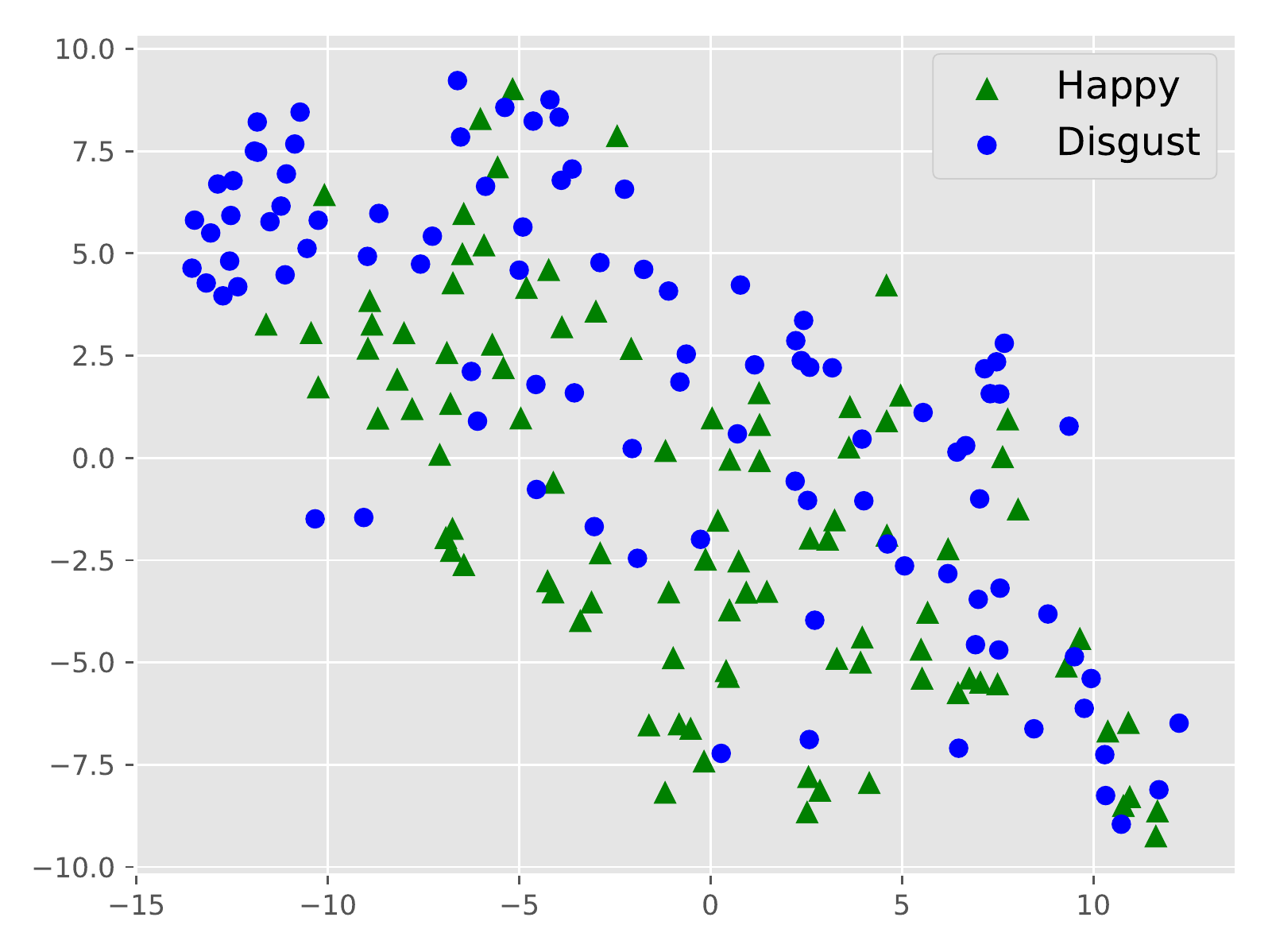}
        \caption{Phoneme Emebdding}
        \label{fig:tsne_phone}
    \end{subfigure}
    \begin{subfigure}[b]{0.33\textwidth}
        \includegraphics[width=\textwidth]{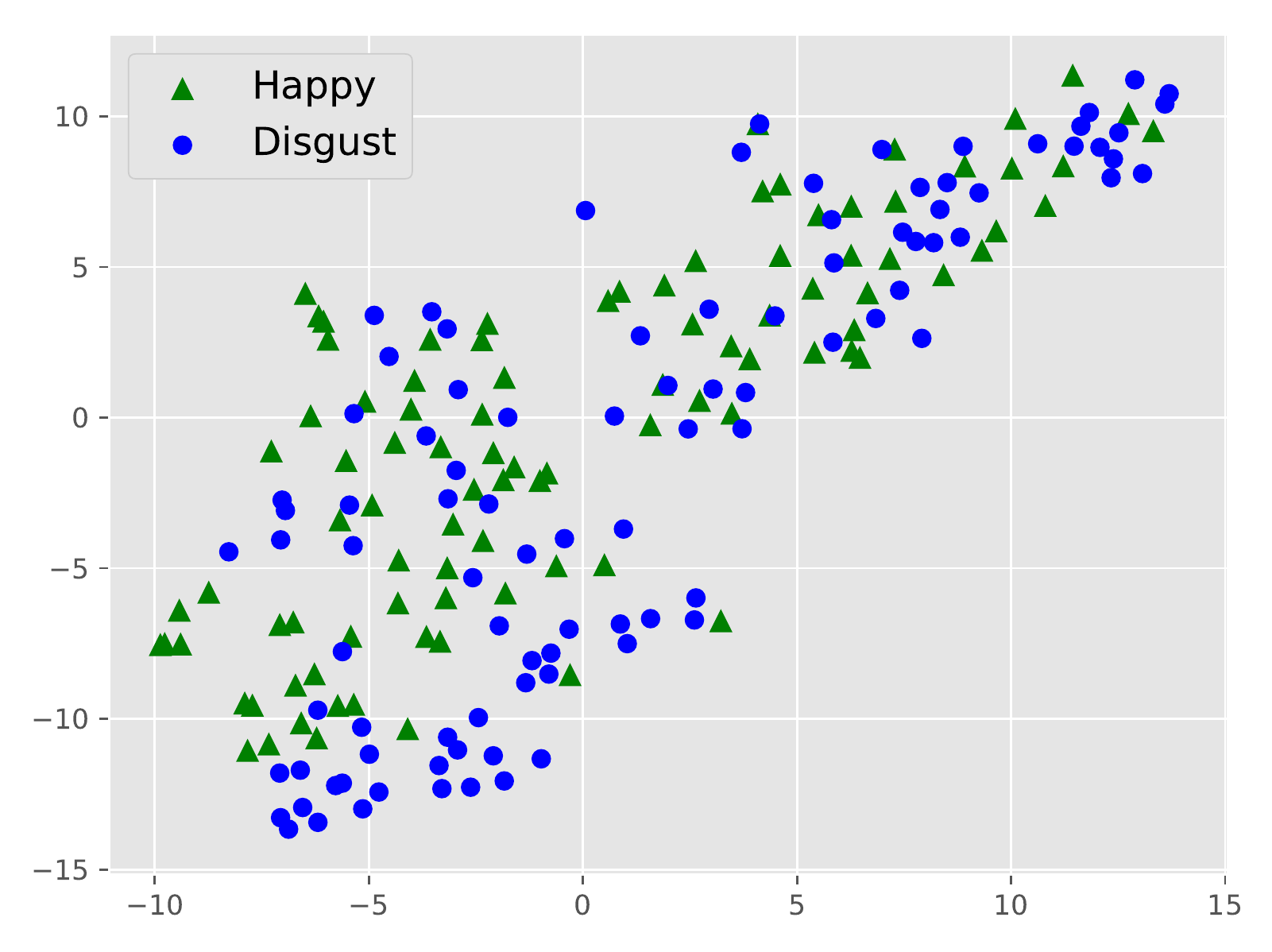}
        \caption{Word Embedding}
        \label{fig:tsne_word}
    \end{subfigure}
    \begin{subfigure}[b]{0.33\textwidth}
        \includegraphics[width=\textwidth]{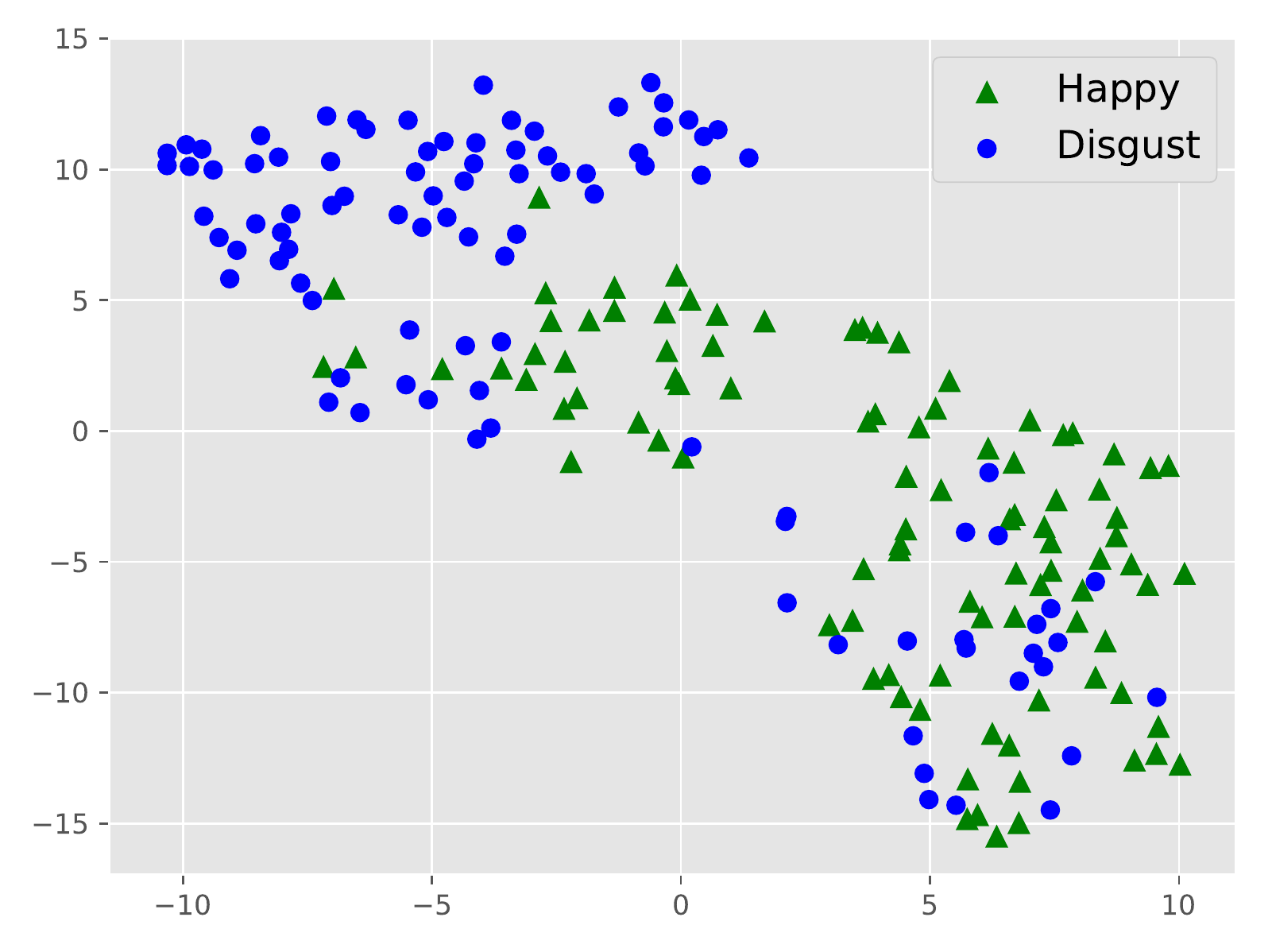}
        \caption{Spoken Sentence Embedding (Ours)}
        \label{fig:tsne_sentence}
    \end{subfigure}
    \caption{\textbf{Visualization of how our method can separate emotions.} Each point denotes an audio clip. The $x$ and $y$ axes denote T-SNE projection axes \cite{maaten2008visualizing}. First, the model was trained on LibriSpeech for speech recognition. Second, the model was used as a fixed feature extractor on the RAVDESS emotion dataset. Finally, the features were plotted and colored by label. }\label{fig:tsne}
\end{figure*}

\subsubsection{Fusion Method}
For these baselines, once we have embeddings for each word or phoneme, we must combine them into a single sentence embedding.
We evaluate two fusion methods:
\begin{enumerate}[itemsep=0pt,topsep=4pt,leftmargin=12pt]
    \item \textit{Uniform Average} \cite{cer2018universal}. The intermediate embeddings are converted to a sentence embedding by computing an element-wise sum of the intermediate embeddings at each word or phoneme position \cite{cer2018universal} and dividing by the number of words in the sentence. This is a uniform average.
    \item \textit{Deep Averaging Network (DAN)} \cite{iyyer2015deep}. First, a uniform average is computed over the intermediate embeddings. Then, the averaged vector is fed into a deep neural network to produce the final sentence embedding. This neural network is trained with the encoder, using the same multitask loss objective from Section \ref{sec:multitask}.
\end{enumerate}
For DAN, a deep neural network converts the simple average into a sentence embedding.
This neural network is the encoder component of a larger encoder-decoder model \cite{sutskever2014sequence} which is trained offline.
This encoder-decoder model is formulated as a multitask learning problem with two decoders: one for acoustic and one for semantic (or linguistic) content.

\section{Results}
\subsection{Speech Processing Tasks}

After training our model on LibriSpeech, we take the encoder and use that to generate spoken sentence embeddings on TIMIT \cite{timit} and RAVDESS \cite{livingstone2018ryerson}.
The weights of encoder are fixed after training and then is only used to extract embeddings.
We now have sentence embeddings for TIMIT and RAVDESS and can proceed to train a simple RNN decoder for automatic speech recognition and a SVM classifier for emotion recognition.
Table \ref{tab:sota} shows results for both tasks.

Sentence-level embeddings outperform phoneme and word embeddings on both tasks.
This is true for both the RNN and our fully convolutional TCN model.
Phoneme-level embeddings perform poorly for speech recognition (WER in excess of 70\%) in contrast to the embedding learned by our method. 
Comparing the fusion methods, the deep averaging network \cite{iyyer2015deep} did not significantly outperform uniform average.
The uniformly averaged vector (i.e., input) was already the information bottleneck.

\subsection{Ablation Studies}

\textbf{Visualization of the Embedding Space.} 
For this experiment, we took our pre-trained spoken sentence encoder, trained on LibriSpeech, and extracted sentence embeddings for each RAVDESS \cite{livingstone2018ryerson} audio clip.
We then visualized the embeddings for different emotions in Figure \ref{fig:tsne}.
Note that our model never saw a training example from RAVDESS.
Without being trained on emotion recognition, our sentence embedding can still cluster emotions. 

\begin{figure}[t]
\vspace{-8mm}
\includegraphics[width=1.0\linewidth]{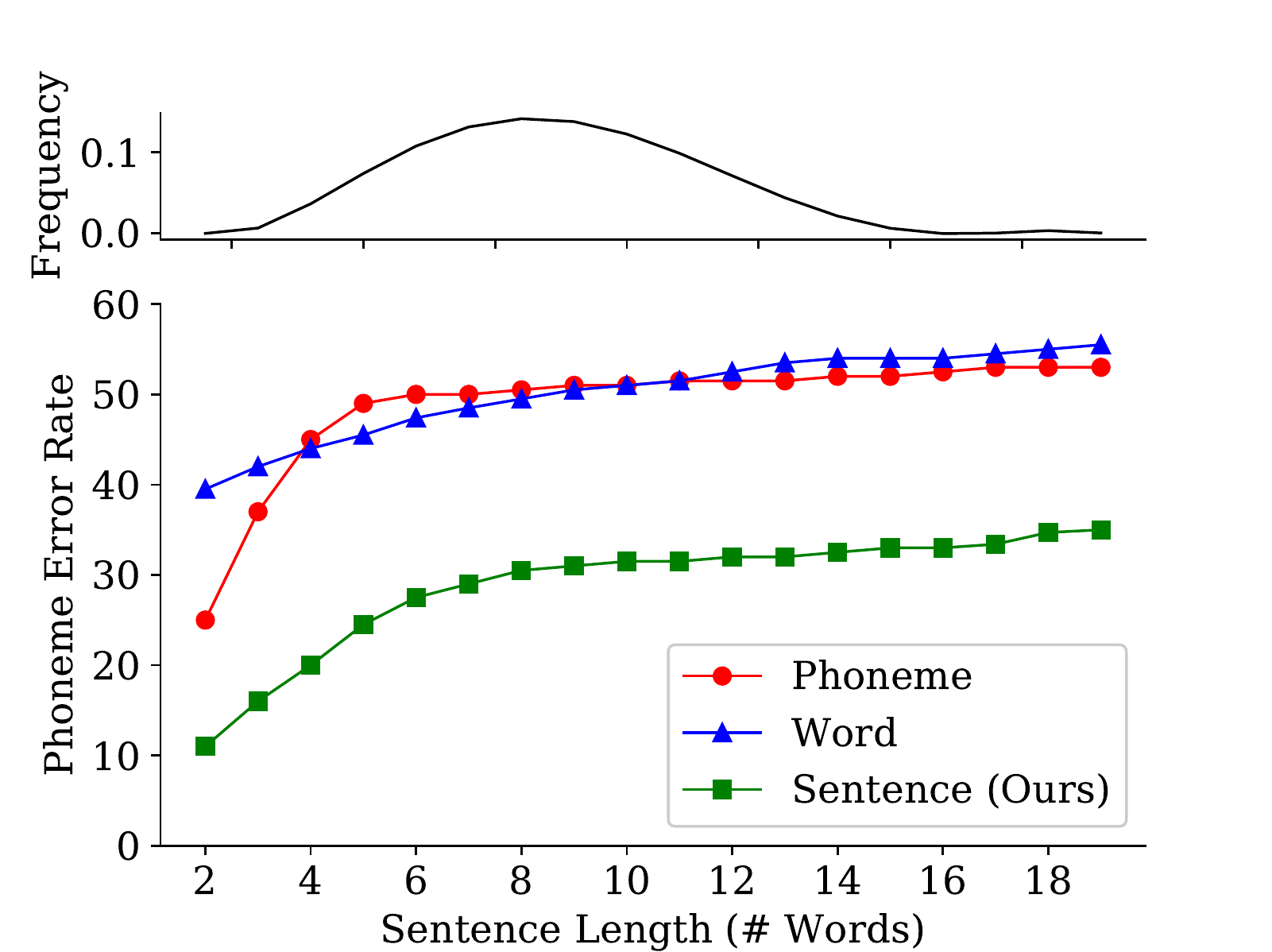}
\caption{\textbf{Effect of sequence length on TIMIT.} (Top) Distribution of sentence lengths. (Bottom) Speech recognition phoneme error rate for different embeddings. }
\label{fig:length}
\end{figure}

\textbf{Sequence Length vs Performance.}
Figure \ref{fig:length} shows the effect of sequence length on speech recognition performance.
In general, as the sentence becomes longer, the performance of the simple embeddings demonstrate lower performance. 
Our method also demonstrates such a pattern (albeit, to a lesser extent), because longer sentences have more content to ``fit" into the fixed-size embedding.
This problem could be remedied with variable-length embeddings, which grow linearly in size with the length of the sentence.

\section{Discussion}

\textbf{Related Work.}
The concept of sentence representations (and longer \cite{le2014distributed}) has been explored in both the natural and spoken language processing communities.
On the natural language side, skip-thought vectors \cite{kiros2015skip} and paraphrastic sentence embeddings \cite{wieting2015towards} have been proposed.
On the speech side, whole sentence maximum entropy models \cite{chen1999efficient}, whole sentence models \cite{huang2018whole}, and time delay networks \cite{peddinti2015time} have been proposed.
Closely related to our work is Speech2Vec \cite{chung2018speech2vec}, which was used in our experiments.
While Speech2Vec is designed for words, it could be extended to sentences as well.

Most similar to our work is the Universal Sentence Encoder (USE) from natural language processing \cite{cer2018universal}.
They propose two sentence encoders.
Each encoder accepts a sentence and produces a single vector.
The first encoder is based on attention \cite{vaswani2017attention}.
The second encoder averages individual word embeddings and feeds the result into a neural network \cite{iyyer2015deep}.
In contrast to USE, our method does not model intermediate word-level features but instead directly learns a sentence embedding.
We believe this can better model long-term context as inter-word relationships are not lost through averaging.

\textbf{Conclusion \& Future Work}.
In this work, we presented a method for learning spoken sentence embeddings which capture both acoustic and linguistic content.
We formulated the problem as a multi-task learning problem to reconstruct both acoustic and linguistic information.
Our results show that our spoken sentence embeddings can be used for emotion recognition and speech recognition.
Future work can focus on learning ``universal" embeddings for spoken language.
There has been promising work on the natural language side \cite{cer2018universal}, of which speech and audio can provide additional dimensions.
Similar to document-level embeddings for text \cite{le2014distributed}, further audio research could explore larger temporal context sizes for full-length songs and multi-sentence recordings. 
Overall, our work illustrates the viability of generic, multi-modal sentence embeddings for spoken language understanding.

\clearpage
\newpage
\small
\bibliographystyle{IEEEbib}
\bibliography{references}

\end{document}